\documentclass[proceedings]{JHEP3}
\usepackage{eurosym}
\usepackage{amssymb}
\usepackage{amsfonts}
\usepackage{amsmath,epsfig}
\usepackage{graphicx}

\newbox\mybox

\newcommand\fverb{\setbox\mybox=\hbox\bgroup\verb}
\newcommand\fverbdo{\egroup\medskip\noindent\fbox{\unhbox\mybox}\ }
\newcommand\fverbit{\egroup\item[\fbox{\unhbox\mybox}]}
\conference{Complex BPS solitons with real energies from duality}
\abstract{Following a generic approach that leads to Bogomolny-Prasad-Sommerfield (BPS) soliton solutions by imposing self-duality,
we investigate three different types of non-Hermitian field theories. We consider a complex version of a logarithmic potential that
possess BPS super-exponential kink and antikink solutions and two different types of complex generalisations of systems of coupled 
sine-Gordon models with kink and antikink solution of complex versions of arctan type. Despite the fact that all soliton solutions obtained
in this manner are complex in the non-Hermitian theories we show that they possess real energies. For the complex extended sine-Gordon model
we establish explicitly that the energies are the same as those in an equivalent pair of a non-Hermitian and Hermitian theory obtained from a pseudo-Hermitian approach 
by means of a Dyson map. We argue that the reality of the energy is due to the topological properties of the complex BPS solutions. These properties
result in general from modified versions of antilinear CPT symmetries that relate self-dual and an anti-self-dual theories.}

\title{Complex BPS solitons with real energies from duality}
\author{Andreas Fring and Takanobu Taira \\
Department of Mathematics, City, University of London,\\
Northampton Square, London EC1V 0HB, UK \\
E-mail: a.fring@city.ac.uk, takanobu.taira@city.ac.uk}

\input{tcilatex}
\begin{document}

\section{Introduction}

T'Hooft \cite{t1974magnetic} and Polyakov \cite{polyakov1974particle}
established more than 45 years ago that gauge theories almost inevitably
contain monopole solutions. The corresponding soliton solutions that
interpolate between different vacua of the theory are usually constructed
explicitly by means of Bogomolny-Prasad-Sommerfield (BPS) \cite%
{bogomol1976stability,prasad1975exact} multiple scaling limits. The validity
of these limits can be justified on physical grounds when assuming that
certain mass ratios in the theory are very small. We have recently
demonstrated \cite{fring2020t} that such type of solutions can also be
constructed in certain domains of the parameter space of a non-Hermitian
field theory with local non-Abelian $SU(2)$ gauge symmetry and a modified
antilinear $\mathcal{CPT}$-symmetry.

Here we investigate the properties of complex soliton solutions resulting in
a general setting of BPS theories and in particular show that the reality of
their energies are attributed to a modified version of a $\mathcal{CPT}$%
-symmetry that relates a self-dual to an anti-self-dual theory and governs
their topological properties. The underlying reason that ensures the reality
of the energy is slightly different to what has been observed previously for
integrable complex nonlinear equations, such as for instance complex
versions of the Korteweg-de Vries equation or Calogero-Sutherland-Moser
systems \cite{AF,AFKdV,PEGAAFint,Mon1,CFB,fring2013pt,CenFring,cen2016time}.
The governing equations to be solved have additional specific structures
that need to be respected by the $\mathcal{CPT}$-symmetry. First of all, the
soliton solutions solve the BPS equations, that are by construction of lower
order than the equations of motion. Moreover, while the soliton solutions
studied in \cite{CFB,CenFring,cen2016time} all vanish asymptotically, the
soliton solutions that maybe associated to magnetic monopoles are of kink or
antikink type with nontrivial asymptotic behaviour. It is this latter
topological behaviour that completely governs the energy, which is bounded
from below by the topological charge of the theory, the Bogomolny bound.
Crucially in our approach is that the BPS equations occur in pairs involving
self-dual and anti-self-dual functions of the fields their first order
derivatives that have the same energy. The modified versions of the $%
\mathcal{CPT}$-symmetry relate the solutions of these two pairs of equations.

One may approach the study of BPS systems in several alternative ways. The
original and most direct way is to investigate a concrete full-fledged gauge
theory and carry out the appropriate limits, see \cite{fring2020t} for a
non-Hermitian system. Alternatively one can take the above mentioned general
properties as the defining relations for a BPS theory and derive them in a
simpler setting as was shown for instance in \cite%
{adam2013some,ferreira2019some,adam2019solvable,klimas2019further}. While
the discussion in \cite{adam2013some} is generic for any dimension we
restrict our considerations here to complex scalar field theories in two
dimensions described by Lagrangians of the general form%
\begin{equation}
\mathcal{L=}\frac{1}{2}\eta _{ab}\partial _{\mu }\phi _{a}\partial ^{\mu
}\phi _{b}-\mathcal{V}(\phi ),
\end{equation}%
where $\eta _{ab}$ is a target space metric, the metric $g$ in space-time is
taken to be Lorentzian $\limfunc{diag}g=(1,-1)$ and the potential $V(\phi )$
depends on the complex scalar field components $\phi _{a}$, $a=1,\ldots
,\ell $.

We present three different types of systems purposely chosen to illustrate
different types of features. For all models we construct the explicit
complex BPS kink and antikink soliton solutions, we identify the different
versions of the modified antilinear $\mathcal{CPT}$-symmetry that can be
used to argue that the corresponding energy is real in certain regimes of
the parameter space. The models exhibit different types of symmetry breaking
and appear to possess exceptional points in their energy spectrum. However,
we demonstrate that none of these points is a genuine exceptional point \cite%
{heiss2012physics} in the standard sense of non-Hermitian theories \cite%
{ptbook,Alirev}. We study the stability of the vacua and identify the
explicit soliton solutions that interpolate between them. Our approach to
analyse directly the non-Hermitian system is justified further in section 5
where we present an explicit example of a pair of a non-Hermitian and a
Hermitian Hamiltonian that are related by a nontrivial Dyson map and show
that the energy of the two systems is identical in the well-defined $%
\mathcal{CPT}$-symmetric regime of the parameter space.

Our manuscript is organized as follows: In section 2 we present the general
set up for the study of BPS solitons from the requirement that the theory
contains self-dual or anti-self-dual functionals of the fields and their
derivatives. In section 3, 4 and 5 we investigate three different types of
models in the way described above and in section 6 we state our conclusions.

\section{BPS solitons from self-duality and anti-self-duality}

The authors in \cite{adam2013some} take an energy functional $E$ and a
topological charge $Q$ of the form%
\begin{equation}
E=\frac{1}{2}\int d^{2}x\left( A_{\alpha }^{2}+\tilde{A}_{\alpha
}^{2}\right) ,\quad \text{and\quad }Q=\int d^{2}xA_{\alpha }\tilde{A}%
_{\alpha },  \label{EQ}
\end{equation}%
as a starting point for the setup of a BPS theory, where the quantities $%
A_{\alpha }(\phi ,\partial _{\mu }\phi )$, $\tilde{A}_{\alpha }(\phi
,\partial _{\mu }\phi )$ are functions of the fields $\phi $ appearing in
the Lagrangian $\mathcal{L}$ of the field theory under consideration and at
most of first order derivatives thereof. It is clear that the relations in (%
\ref{EQ}) ensure that the topological charge is always a lower bound for the
energy $E\geq \left\vert Q\right\vert $. Following \cite{adam2013some}, one
may then use these definitions to derive two equations, one being the
Euler-Lagrange equation resulting from varying $E$ and the other from
considering infinitessimal changes $\delta \phi $ in $Q$ and demanding $%
\delta Q=0$. The latter requirement incorporates that $Q$ is interpreted as
a topological charge, which should be a homotopy invariant, i.e. invariant
under smooth variations in the fields. The compatibility between these two
equations then implies (anti)-self-duality of the quantities $A_{\alpha }$, $%
\tilde{A}_{\alpha }$ and moreover that $Q$ saturates the Bogomolny bound for
the energy $E$%
\begin{equation}
A_{\alpha }=\pm \tilde{A}_{\alpha },\quad \text{and\quad }E=\left\vert
Q\right\vert .  \label{self}
\end{equation}%
Evidently the energies of the self-dual and anti-self-dual fields are the
same. Assuming next the existence of a pre-potential $U(\phi )$, that is a
function of the fields in the theory only, one may write\ the energy
functional and the topological charge for the static solutions as%
\begin{eqnarray}
E &=&\int\nolimits_{-\infty }^{\infty }dx\left( \frac{1}{2}\eta
_{ab}\partial _{\mu }\phi _{a}\partial ^{\mu }\phi _{b}+\mathcal{V}(\phi
)\right) =\frac{1}{2}\int\nolimits_{-\infty }^{\infty }dx\left( \eta
_{ab}\partial _{\mu }\phi _{a}\partial ^{\mu }\phi _{b}+\eta _{ab}^{-1}\frac{%
\partial U}{\partial \phi _{a}}\frac{\partial U}{\partial \phi _{b}}\right)
,~~~~~  \label{E2} \\
Q &=&\int\nolimits_{-\infty }^{\infty }dx\frac{\partial U}{\partial x}%
=\int\nolimits_{-\infty }^{\infty }dx\frac{\partial U}{\partial \phi _{a}}%
\partial _{x}\phi _{a}=\lim_{x\rightarrow \infty }U[\phi
(x)]-\lim_{x\rightarrow -\infty }U[\phi (x)].  \label{Q2}
\end{eqnarray}%
Comparing the general expressions for $A_{\alpha }$ and $\tilde{A}_{\alpha }$
in (\ref{EQ}) with those for $U(\phi )$ in (\ref{E2}), (\ref{Q2}) implies
the identifications%
\begin{equation}
A_{a}=\rho _{ab}\partial _{x}\phi _{b},\quad \text{and\quad }\tilde{A}_{a}=%
\frac{\partial U}{\partial \phi _{b}}\rho _{ba}^{-1},
\end{equation}%
where $\rho $ factorizes the target space metric as $\rho ^{T}\rho =\eta $.
The (anti)-self-duality relations in (\ref{self}), then become equivalent to
the pair of BPS equations in the form%
\begin{equation}
\partial _{x}\phi _{b}=\pm \eta _{ab}^{-1}\frac{\partial U}{\partial \phi
_{b}}.  \label{BPS}
\end{equation}%
Allowing the scalar fields to be complex and the potential to be
non-Hermitian, the reality of the energy could be guaranteed when the
Hamiltonian is $\mathcal{CPT}$-symmetric satisfying $\mathcal{H}\left[ \phi
(x)\right] =\mathcal{H}^{\dagger }\left[ \phi (-x)\right] $ by employing the
same argument as in \cite{AFKdV}%
\begin{equation}
E=\int\nolimits_{-\infty }^{\infty }dx\mathcal{H}\left[ \phi (x)\right]
=-\int\nolimits_{\infty }^{-\infty }dx\mathcal{H}\left[ \phi (-x)\right]
=\int\nolimits_{-\infty }^{\infty }dx\mathcal{H}^{\dagger }\left[ \phi (x)%
\right] =E^{\ast }.  \label{EE}
\end{equation}%
Since in the scenario considered here the self-duality imposes the kinetic
energy to equal the potential energy, it would suffice therefore to
establish that%
\begin{equation}
\mathcal{V}\left[ \phi _{\pm }(x)\right] =\mathcal{V}^{\dagger }\left[ \phi
_{\pm }(-x)\right] ,\qquad \text{or\qquad }\mathcal{V}\left[ \phi _{\pm }(x)%
\right] =\mathcal{V}^{\dagger }\left[ \phi _{\mp }(-x)\right]  \label{VV}
\end{equation}%
in order to ensure the reality of the energy by means of (\ref{EE}). We have
denoted here by $\phi _{\pm }$ the solutions of (\ref{BPS}) corresponding to
the two options for the sign in (\ref{self}). Evidently it follows from (\ref%
{EQ}) that the energy is the same for either choice. The second option in (%
\ref{VV}) is novel due to the set up involving anti-self-duality and not
available in the standard setting of integrable systems \cite%
{AFKdV,CFB,CenFring,cen2016time}. We shall demonstrate below that actually
this novel option of the $\mathcal{CPT}$-symmetry is the guarand for the
reality of the energy for the systems considered. Evidently, for a direct
analysis it is clear that the energy is real if%
\begin{equation}
\lim_{x\rightarrow \infty }\func{Im}\left\{ U[\phi (x)]\right\}
=\lim_{x\rightarrow -\infty }\func{Im}\left\{ U[\phi (x)]\right\} .
\label{UU}
\end{equation}

We shall now analyse several different theories with concrete choices for
pre-potential that lead to non-Hermitian scalar field theory with an
antilinear symmetry. We shall demonstrate that the first version of the $%
\mathcal{CPT}$-symmetry in (\ref{VV}) is in fact broken, but the second
version can be realised by the various solutions in our examples.

\section{A non-Hermitian BPS theory with super-exponential kink solutions}

We start by generalizing a Hermitian one field theory that was recently
studied by Kumar, Khare and Saxena \cite{kumar2019model} to one with two
component complex fields in a non-Hermitian setting. The original model was
motivated in parts by its proximity to a $\phi ^{6}$-type potential and its
feature of minimal nonlinearity. A very interesting aspect of this model is
that it possesses kink and antikink solutions with a super-exponential
profile rather than the more standard $\arctan $ type solutions. This
feature survives our generalization and moreover the complex BPS solutions
interpolating between five out of nine vacua of our model have real energies.

To set up the field theory we choose the target space metric and the
pre-potential as%
\begin{equation}
\eta =\left( 
\begin{array}{ll}
1 & -i\lambda \\ 
-i\lambda & 1%
\end{array}%
\right) ,\quad \text{and\quad }U\left( \phi _{1},\phi _{2}\right) =\frac{\mu
_{1}}{2}\phi _{1}^{2}\ln \left( \phi _{1}^{2}\right) +\frac{\mu _{2}}{2}\phi
_{2}^{2}\ln \left( \phi _{2}^{2}\right) ,~~~~~\ \lambda ,\mu _{1},\mu
_{2}\in \mathbb{R\,},  \label{an}
\end{equation}%
respectively. Using the relation between the potential and the pre-potential
(\ref{E2}) we obtain from the Ansatz (\ref{an}) the non-Hermitian potential%
\begin{equation}
\mathcal{V}\left( \phi _{1},\phi _{2}\right) =\frac{1}{1+\lambda ^{2}}%
\sum\limits_{i=1}^{2}\frac{\mu _{i}^{2}}{2}\left[ \text{$\phi $}_{i}+\text{$%
\phi $}_{i}\ln \left( \text{$\phi $}_{i}^{2}\right) \right] ^{2}+i\frac{%
\lambda }{1+\lambda ^{2}}\prod\limits_{i=1}^{2}\mu _{i}\left[ \text{$\phi $}%
_{i}+\text{$\phi $}_{i}\ln \left( \text{$\phi $}_{i}^{2}\right) \right] .
\label{V1}
\end{equation}%
According to the standard pseudo-Hermitian approach to non-Hermitian field
theories one may seek a similarity transformation by means of a well defined
Dyson map, e.g. \cite%
{Bender:2005hf,mannheim2018goldstone,fring2020goldstone,fring2019pseudo,fring2020massive}%
, to map the theory to a Hermitian theory or introduce non-vanishing surface
terms \cite{Kings1,Kings2,Kings3,Kings4,Kings5,Kings6} and analyse these
systems. However, as we shall demonstrate below, just as in a standard
quantum mechanical setting \cite{ptbook,Alirev}, the energy is preserved in
this process so that one may also analyse the solutions of the non-Hermitian
theory directly. Our approach is further justified in section 5 where we
shall present an explicit system for which a non-Hermitian Hamiltonian is
related to a Hermitian Hamiltonian by means of an explicit nontrivial Dyson
map.

Using the BPS equations (\ref{BPS}), the static solutions associated to the
potential (\ref{V1}) are the two pairs of coupled first order differential
equations%
\begin{eqnarray}
BPS_{1}^{\pm } &:&\qquad \partial _{x}\phi _{1}=\pm \frac{\mu _{1}\left[ 
\text{$\phi $}_{1}+\text{$\phi $}_{1}\ln \left( \text{$\phi $}%
_{1}^{2}\right) \right] }{\lambda ^{2}+1}\pm i\lambda \frac{\mu _{2}\left[ 
\text{$\phi $}_{2}+\text{$\phi $}_{2}\ln \left( \text{$\phi $}%
_{2}^{2}\right) \right] }{\lambda ^{2}+1}=:F_{1}^{\pm },  \label{B1} \\
BPS_{2}^{\pm } &:&\qquad \partial _{x}\phi _{2}=\pm i\lambda \frac{\mu _{1}%
\left[ \text{$\phi $}_{1}+\text{$\phi $}_{1}\ln \left( \text{$\phi $}%
_{1}^{2}\right) \right] }{\lambda ^{2}+1}\pm \frac{\mu _{2}\left[ \text{$%
\phi $}_{2}+\text{$\phi $}_{2}\ln \left( \text{$\phi $}_{2}^{2}\right) %
\right] }{\lambda ^{2}+1}=:F_{2}^{\pm }.  \label{B2}
\end{eqnarray}%
We will need both versions in (\ref{B1}) and (\ref{B2}) to verify the
general argument that guarantees the reality of the energy. We observe that
these equations are compatible under two types of modified $\mathcal{CPT}$%
-transformations 
\begin{equation}
\mathcal{CPT}_{\pm }:~\phi _{1}(x)\rightarrow \pm \left[ \phi _{1}(-x)\right]
^{\dagger }\text{, ~~\ ~}\phi _{2}(x)\rightarrow \mp \left[ \phi _{2}(-x)%
\right] ^{\dagger },\text{\quad }\Leftrightarrow \quad BPS_{i}^{\pm
}\rightarrow \left( BPS_{i}^{\mp }\right) ^{\ast }.
\end{equation}%
Using these symmetries we can derive the second relation in (\ref{VV}). We
notice that a modified $\mathcal{CT}$-transformation $\phi
_{1}(x)\rightarrow -\left[ \phi _{1}(x)\right] ^{\dagger }$, $\phi
_{2}(x)\rightarrow -\left[ \phi _{2}(x)\right] ^{\dagger }$ is achieving the
compatibility $BPS_{i}^{\pm }\rightarrow \left( BPS_{i}^{\pm }\right) ^{\ast
}$. However, this symmetry can not be employed in the argument in (\ref{EE})
that guarantees the reality of the energy. The introduction of time by means
of a standard Lorentz transformation, $x\rightarrow (x-vt)/\sqrt{1-v^{2}}$,
will not change this feature, so that the reality of the energy is not a
consequence of this particular antilinear symmetry. Moreover, we do not find
solutions below that posses this kind of $\mathcal{CT}$-symmetry.

Let us now solve the pair of the two BPS equations (\ref{B1}) and (\ref{B2}%
). In the Hermitian limit, when $\lambda =0$, the equations decouple and the
solutions can be obtained in an explicit analytical form as double
exponentials%
\begin{equation}
\phi _{i}(x)=\exp \left( -\frac{1}{2}+\frac{1}{2}e^{2(\mu _{i}x+\kappa
_{i})}\right) ,
\end{equation}%
with integration constants $\kappa _{i}\in \mathbb{C}$ and $i=1,2$. We fix
our constants in such a way that we obtain proper kink and antikink
solutions with well-defined asymptotic behaviour. We select our solutions as%
\begin{eqnarray}
\phi _{i}^{a+}(x) &=&\exp \left( -\frac{1}{2}-\frac{1}{2}e^{2\mu
_{i}x}\right) ,~\ \ \ \phi _{i}^{k+}(x)=-\exp \left( -\frac{1}{2}-\frac{1}{2}%
e^{2\mu _{i}x}\right) ,~~~\mu _{i}\geq 0, \\
\phi _{i}^{k-}(x) &=&\exp \left( -\frac{1}{2}-\frac{1}{2}e^{2\mu
_{i}x}\right) ,~\ \ \ \phi _{i}^{a-}(x)=-\exp \left( -\frac{1}{2}-\frac{1}{2}%
e^{2\mu _{i}x}\right) ,~~~\ \mu _{i}<0,
\end{eqnarray}%
so that $\phi _{i}^{a+}(0)=\phi _{i}^{k-}(0)=\allowbreak 1/e$, $\phi
_{i}^{k+}(0)=\phi _{i}^{a-}(0)=-1/e$ and $\phi _{i}^{a+}(x)=\phi
_{i}^{k-}(-x)=-\phi _{i}^{k+}(x)=-\phi _{i}^{a-}(-x)$. The asymptotic limits
are therefore%
\begin{eqnarray}
\lim_{x\rightarrow -\infty }\phi _{i}^{a+}(x) &=&\lim_{x\rightarrow \infty
}\phi _{i}^{k-}(x)=\frac{1}{\sqrt{e}}\text{, ~\ \ ~}\lim_{x\rightarrow
-\infty }\phi _{i}^{k+}(x)=\lim_{x\rightarrow \infty }\phi _{i}^{a-}(x)=-%
\frac{1}{\sqrt{e}},\text{\ \ \ } \\
\text{\ }\lim_{x\rightarrow \infty }\phi _{i}^{a+}(x) &=&\lim_{x\rightarrow
\infty }\phi _{i}^{k+}(x)=\lim_{x\rightarrow -\infty }\phi
_{i}^{a-}(x)=\lim_{x\rightarrow -\infty }\phi _{i}^{k-}(x)=0\text{. }
\end{eqnarray}%
Hence, using the expression for the pre-potential (\ref{Q2}) we obtain for
all combinations the same real energy as function of $\mu _{1}$, $\mu _{2}$%
\begin{equation}
E^{\phi _{1}^{pn},\phi _{2}^{qm}}(\mu _{1},\mu _{2})=\frac{\left\vert \mu
_{1}\right\vert +\left\vert \mu _{2}\right\vert }{2e},~~\ \ \
p,q=k,a;~n,m=\pm ;~~~\mu _{1},\mu _{2}\in \mathbb{R}.  \label{EM}
\end{equation}%
In the non-Hermitian scenario, when $\lambda \neq 0$, we solve the two sets
of coupled BPS equations (\ref{B1}) and (\ref{B2}) numerically. Some sample
computations are presented in figure \ref{Fig0}.

\FIGURE{ \epsfig{file=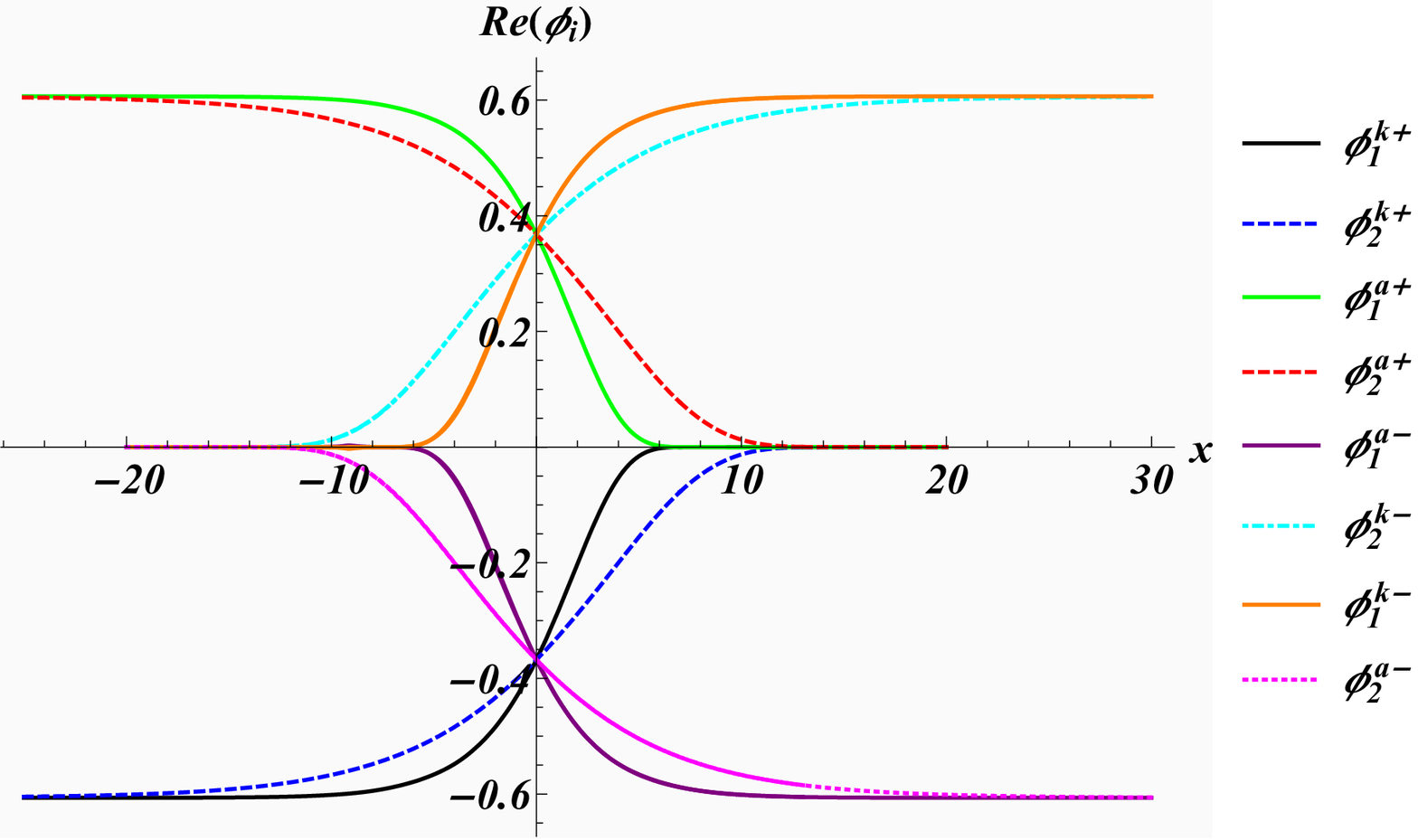, width=7.7cm} \epsfig{file=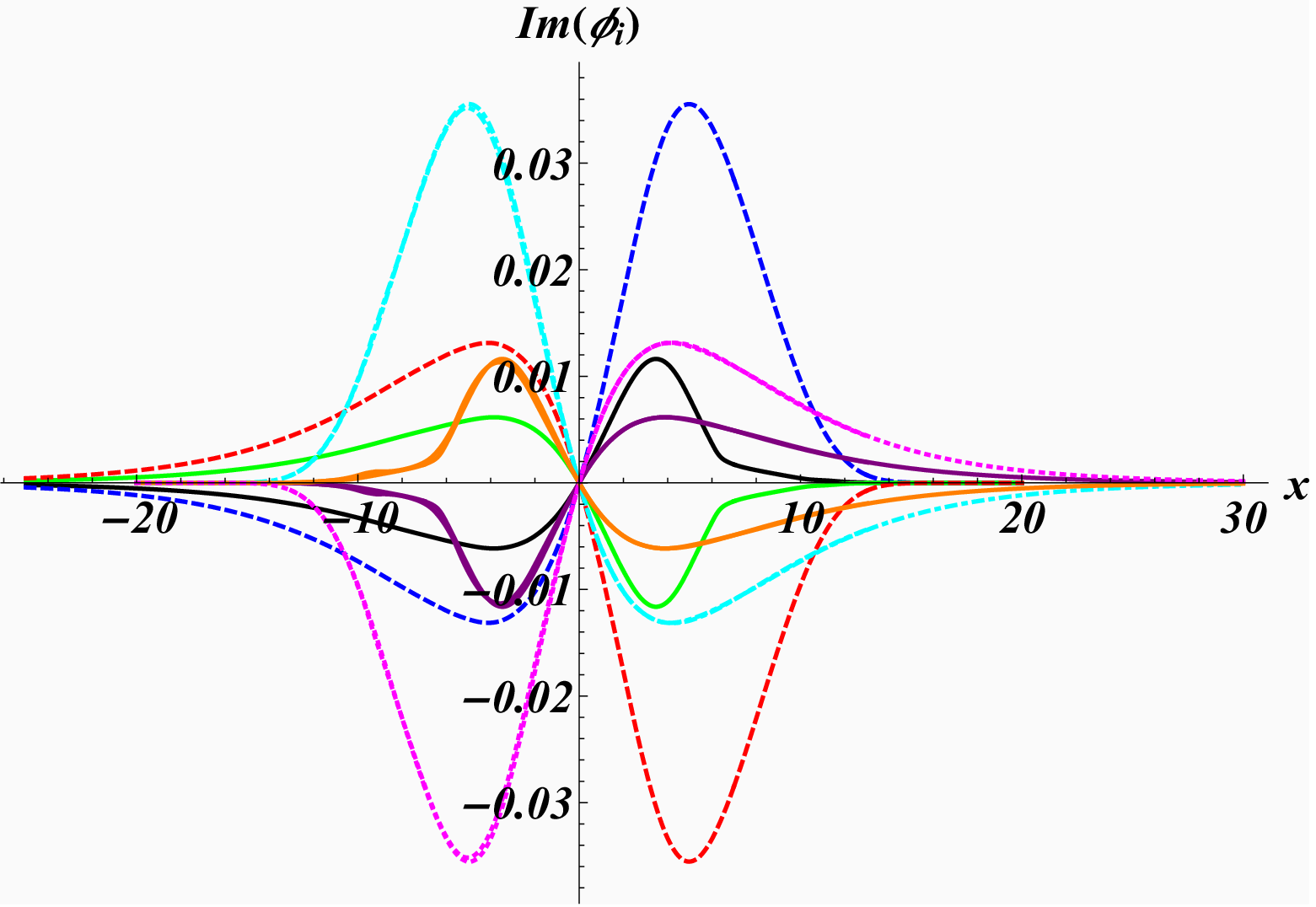,width=6.7cm}
\caption{Complex BPS kink and anitkink solutions of the two pairs of coupled BPS equations (\ref{B1}) and (\ref{B2}) associated to the potential (\ref{V1}) with initial 
values $\phi _{1}^{k+}(0)=\phi
_{2}^{k+}(0)=\phi _{1}^{a-}(0)=\phi _{2}^{a-}(0)=-1/e$ and $\phi
_{1}^{a+}(0)=\phi _{2}^{a+}(0)=\phi _{1}^{k-}(0)=\phi _{2}^{k-}(0)=1/e$
 for $\mu _{1}=0.2$, $\mu _{2}=0.1$,  $\lambda = 0.1 $. }
       \label{Fig0}}

We observe that for increasing values of the coupling constants $\mu _{i}$
the real parts of $\phi _{i}$ approach $H(-x)/\sqrt{e}\ $with $H(x)$
denoting the Heaviside step function. The imaginary parts keep oscillating
with increased amplitudes and crucially vanish at $x\rightarrow \pm \infty $%
, which means that the energy is given by the expression in (\ref{EM}) for
all values of $\lambda $.

We also observe from our numerical solutions in figure \ref{Fig0} that the
solutions realize the $\mathcal{CPT}_{-}$-symmetry as 
\begin{equation}
\phi _{1}^{k+}(x)=-\left[ \phi _{1}^{k-}(-x)\right] ^{\dagger },~~~~\phi
_{1}^{a+}(x)=-\left[ \phi _{1}^{a-}(-x)\right] ^{\dagger },~~~~\phi
_{2}^{k\pm }(x)=\left[ \phi _{2}^{a\mp }(-x)\right] ^{\dagger }.
\label{prop}
\end{equation}%
Using now the properties of the kink and antikink solutions (\ref{prop}) we
derive for the potential%
\begin{equation}
\mathcal{V}_{\lambda }\left[ \phi _{1}^{k+}(x),\phi _{2}^{k+}(x)\right] =%
\mathcal{V}_{\lambda }\left\{ -\left[ \phi _{1}^{k-}(-x)\right] ^{\dagger },%
\left[ \phi _{2}^{a-}(-x)\right] ^{\dagger }\right\} =\mathcal{V}_{\lambda
}^{\dagger }\left\{ \left[ \phi _{1}^{k-}(-x)\right] ,\left[ \phi
_{2}^{a-}(-x)\right] \right\} ,  \label{Ve}
\end{equation}%
and similarly for the others pairs of solutions. Changing the initial
conditions we may also construct solutions that manifest the $\mathcal{CPT}%
_{+}$-symmetry. The relation in (\ref{Ve}) is precisely the second option in
(\ref{VV}) that relates solutions of the self-dual system to solutions of
the anti-self-dual system. As the energies in both systems must be the same
it is guaranteed to be real.

Next we will identify which vacua are interpolated by which kind of BPS
solution. It is easy to check that the real part of the potential has nine
minima at 
\begin{equation}
v^{\pm \pm }=(\pm e^{-1/2},\pm e^{-1/2}),~~v^{0\pm }=(0,\pm
e^{-1/2}),~~v^{\pm 0}=(\pm e^{-1/2},0),~~v^{00}=(0,0),
\end{equation}%
corresponding to the fixed points of the dynamical system (\ref{B1}) and (%
\ref{B2}) as solutions of $F_{1}^{\pm }(\phi _{1},\phi _{2})=F_{2}^{\pm
}(\phi _{1},\phi _{2})=0$. Next we compute the eigenvalues of the Jacobian
matrix at these fixed points%
\begin{equation}
J=\left. \left( 
\begin{array}{ll}
\partial _{\phi _{1}}F_{1}^{\pm } & \partial _{\phi _{2}}F_{1}^{\pm } \\ 
\partial _{\phi _{1}}F_{2}^{\pm } & \partial _{\phi _{2}}F_{2}^{\pm }%
\end{array}%
\right) \right\vert _{v^{i,j}},~~~~\ \ i,j=0,+,-,
\end{equation}%
in order to determine their stability. For the $F^{+}$-system with $\mu
_{i}>0$ we find that $J\left( v^{\pm \pm }\right) $ has two positive
eigenvalues, $J\left( v^{00}\right) $ has two negative eigenvalues and $%
J\left( v^{0\pm }\right) $, $J\left( v^{\pm 0}\right) $ have a positive and
a negative eigenvalue. For $\phi _{i}\rightarrow 0$ we have to evaluate the
values in an $\varepsilon $-neighbourhood. This means, see e.g. \cite%
{arrowsmith92}, that $v^{\pm \pm }$ are unstable fixed points, $v^{0\pm }$
and $v^{\pm 0}$ are saddle points and $v^{00}$ is the only stable fixed
point. For the $F^{-}$-system still with $\mu _{i}>0$ all signs of the
eigenvalues are reversed. Changing the sign of $\mu _{i}$ will also reverse
the sign of one eigenvalue. Using the solutions from above as represented in
figure \ref{Fig0}, we have the following interpolations between the
different vacua 
\begin{equation}
v^{--}~\underrightarrow{\phi _{1}^{k+}\phi _{2}^{k+}}~v^{00},~~~v^{00}~%
\underrightarrow{\phi _{1}^{a-}\phi _{2}^{k-}}~v^{-+},~~~v^{00}~~%
\underrightarrow{\phi _{1}^{k-}\phi _{2}^{a-}}~v^{+-},~~~v^{++}~~%
\underrightarrow{\phi _{1}^{a+}\phi _{2}^{a+}}~v^{00}.  \label{inter}
\end{equation}%
This behaviour is also confirmed by the gradient flow for $F^{+}$ that is
indicated in figure \ref{Fig01} superimposed onto the potential. We obtain
similar relations for the $F^{-}$-system.

\FIGURE{ \epsfig{file=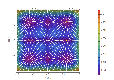, width=10cm} 
\caption{Real part of the potential $\mathcal{V}\left( \phi
_{1},\phi _{2}\right) $ in (\ref{V1}) as a function of $\func{Re}\phi _{1}$ and $\func{Re}\phi _{2}$ with the gradient flow of the real parts of $F^+$ superimposed in white.
The kink-kink, kink-antikink, antikink-kink
and antikink-antikink interpolate between the different types of stable and unstable vacua as specified in (\ref{inter})}
       \label{Fig01}}

When passing from the $\mathcal{V}_{+}$-theory to the $\mathcal{V}_{-}$%
-theory we pass through the special point $\mu _{1}=\mu _{2}=0$. The energy (%
\ref{EM}) is defined for all values and does not become complex. To
investigate this point further the next model is designed in such a way that
it appears to have an exceptional point, which, however, turns out to be not
genuine.

\section{A non-Hermitian coupled sine-Gordon model}

Next we consider a modified version of a model whose real variant has been
investigated recently in \cite{ferreira2019some}. We generalize that model
to one involving a complex non-Hermitian potential with a complex
two-component scalar field and add an additional term designed in such a way
that we apparently obtain an exceptional point \cite{heiss2012physics}. We
shall demonstrate that the system possesses complex solutions to its BPS
equations with real energies in a certain region in the parameter space
where the topological charge of the system is well-defined and real. There
is also a region in which the energy is not well defined and not finite on
the entire real $x$-axis.

Choosing the target space metric and the pre-potential as%
\begin{equation}
\eta =\left( 
\begin{array}{ll}
1 & -i\lambda  \\ 
-i\lambda  & 1%
\end{array}%
\right) ,\quad \text{and\quad }U\left( \phi _{1},\phi _{2}\right) =-(\cos
\phi _{1}+\mu \phi _{1}+\cos \phi _{2}),~~\ \lambda ,\mu \in \mathbb{R\,},
\end{equation}%
respectively, the potential resulting from the expression in (\ref{E2}) is
derived as%
\begin{equation}
\mathcal{V}\left( \phi _{1},\phi _{2}\right) =\frac{1}{2(1+\lambda ^{2})}%
\left[ \left( \sin \phi _{1}-\mu \right) ^{2}+2i\lambda \left( \sin \phi
_{1}-\mu \right) \sin \phi _{2}+\sin ^{2}\phi _{2}\right] .  \label{V}
\end{equation}%
We note that the singularity at $\lambda =1$ present in the real version of
this model discussed in \cite{ferreira2019some} has been removed. The static
versions of the BPS equations (\ref{BPS}) obtained from (\ref{V}) are the
pairs of complex coupled first order equations%
\begin{eqnarray}
BPS_{1}^{\pm } &:&\qquad \partial _{x}\phi _{1}=\pm \frac{1}{1+\lambda ^{2}}%
\left( \sin \phi _{1}-\mu +i\lambda \sin \phi _{2}\right) =:G_{1}^{\pm },
\label{BPS1} \\
BPS_{2}^{\pm } &:&\qquad \partial _{x}\phi _{2}=\pm \frac{1}{1+\lambda ^{2}}%
\left[ i\lambda \left( \sin \phi _{1}-\mu \right) +\sin \phi _{2}\right]
=:G_{2}^{\pm }.  \label{BPS2}
\end{eqnarray}%
These equations are compatible under the modified $\mathcal{CPT}$%
-transformation 
\begin{equation}
\mathcal{CPT}:~\phi _{1}(x)\rightarrow \left[ \phi _{1}(-x)\right] ^{\dagger
}\text{, ~~\ ~}\phi _{2}(x)\rightarrow -\left[ \phi _{2}(-x)\right]
^{\dagger },\text{\quad }\Leftrightarrow \quad BPS_{i}^{\pm }\rightarrow
\left( BPS_{i}^{\mp }\right) ^{\ast }.  \label{A}
\end{equation}%
Notice that we require again both signs to achieve consistency under the $%
\mathcal{CPT}$-conjugation. It is precisely this symmetry that is needed to
derive the second relation in (\ref{VV}). Trying instead to realize the
compatibility of $BPS_{i}^{+}$ or $BPS_{i}^{-}$ with itself requires just a
modified $\mathcal{CT}$ -transformation $\phi _{1}(x)\rightarrow \left[ \phi
_{1}(x)\right] ^{\dagger }$, $\phi _{2}(x)\rightarrow -\left[ \phi _{2}(x)%
\right] ^{\dagger }$, which as for the previous model is, however, not
sufficient to be used in the argument in (\ref{EE}) that ensures the reality
of the energy.

In the Hermitian limit, when $\lambda =0$, the two pairs of BPS equations
decouple and are easily solved by the kink and antikink solutions for the
upper and lower sign, respectively, 
\begin{eqnarray}
\phi _{1}^{\pm (n)}(x) &=&2\arctan \left\{ \frac{1}{^{\mu }}\left[ 1+\sqrt{%
1-\mu ^{2}}\tanh \left[ \frac{1}{2}\sqrt{1-\mu ^{2}}(\pm x+\kappa _{1})%
\right] \right] \right\} +2\pi n,  \label{f1} \\
\phi _{2}^{\pm (n)}(x) &=&2\arctan \left( e^{\pm x+\kappa _{2}}\right) +2\pi
n,  \label{f2}
\end{eqnarray}%
where $n\in \mathbb{Z}$ and integration constants $\kappa _{1},\kappa
_{2}\in \mathbb{R}$. From the asymptotic limits 
\begin{eqnarray}
\lim_{x\rightarrow \infty }\phi _{1}^{+(n)}(x) &=&\lim_{x\rightarrow -\infty
}\phi _{1}^{-(n)}(x)=2n\pi +\func{sign}(\mu )\pi -\arcsin (\mu )\text{, } \\
\lim_{x\rightarrow -\infty }\phi _{1}^{+(n)}(x) &=&\lim_{x\rightarrow \infty
}\phi _{1}^{-(n)}(x)=2n\pi +\func{sign}(\mu )\arcsin (\mu )\text{, ,} \\
\lim_{x\rightarrow \pm \infty }\phi _{2}^{+(n)}(x) &=&\lim_{x\rightarrow \mp
\infty }\phi _{2}^{-(n)}(x)=2n\pi +\frac{\pi \pm \pi }{2}\text{,~}
\end{eqnarray}%
for $\left\vert \mu \right\vert \leq 1$, we obtain from (\ref{Q2}) for both
signs the same expression for the energy as a function of $\mu $%
\begin{equation}
E^{\pm }(\mu )=2\left[ 1+\sqrt{1-\mu ^{2}}-\mu \arctan \left( \frac{\sqrt{%
1-\mu ^{2}}}{\mu }\right) \right] .  \label{Emu}
\end{equation}%
For $\left\vert \mu \right\vert >1$ the limits $\lim_{x\rightarrow \pm
\infty }\phi _{i}(x)$ are not well defined as the solutions become periodic
in this case. Limiting this case to a theory on a finite interval will,
however, still give real energies.

In the non-Hermitian scenario, when $\lambda \neq 0$, we solve the coupled
equations (\ref{BPS1}) and (\ref{BPS2}) numerically, see figure \ref{Fig1}
for some sample behaviours.

\FIGURE{ \epsfig{file=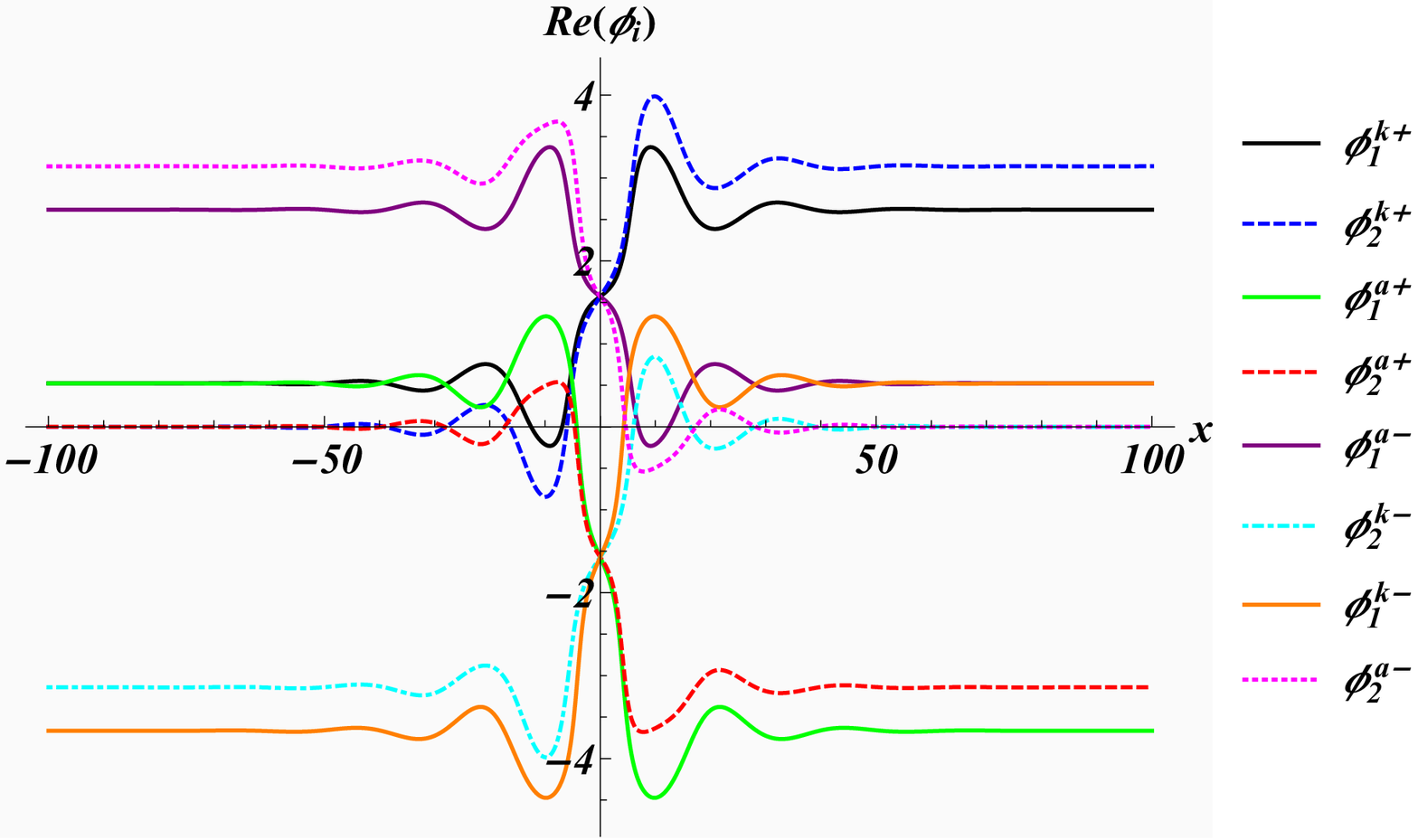, width=7.8cm} \epsfig{file=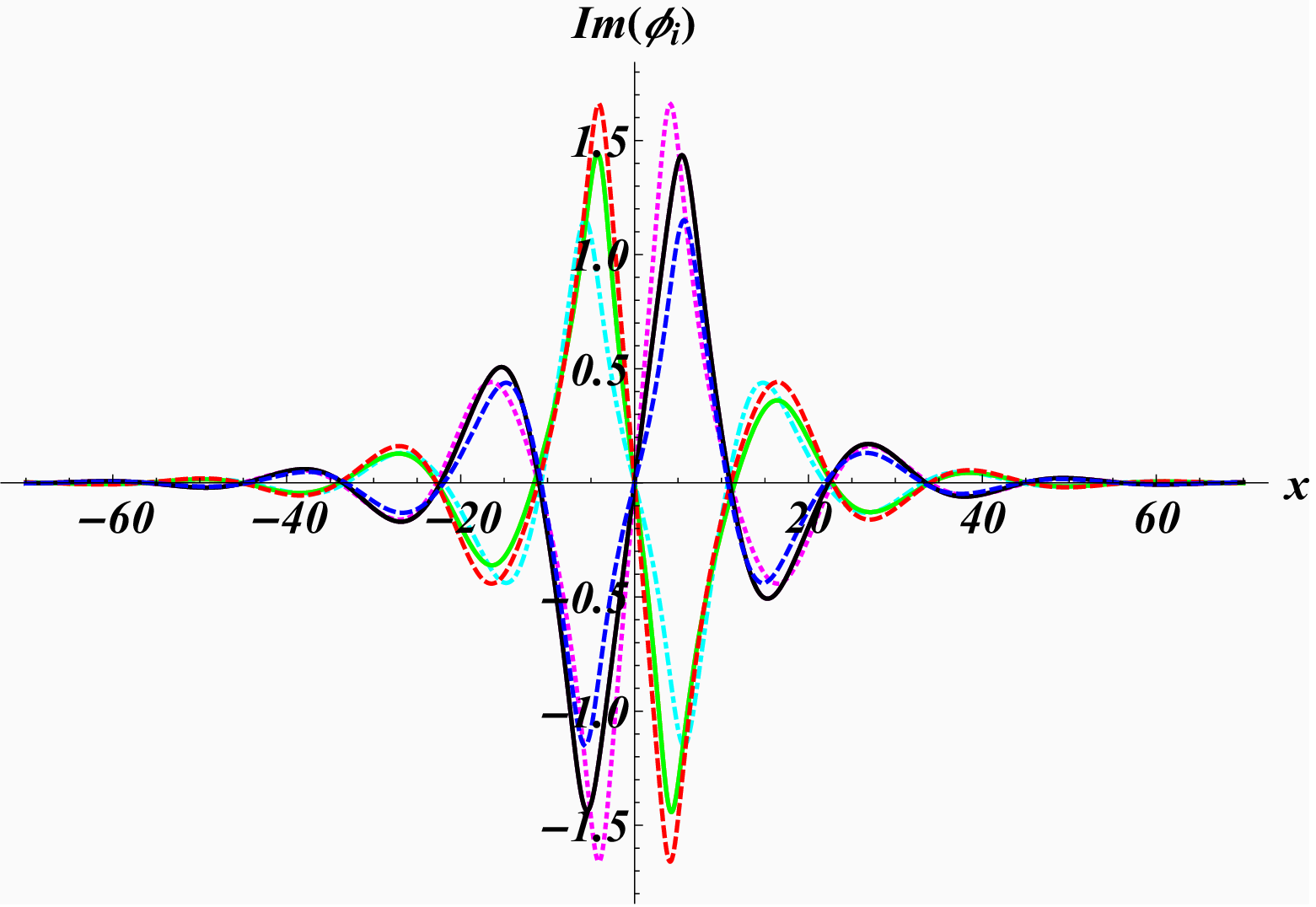,width=6.6cm}
\caption{Complex BPS kink and antikink
solutions of the pair of BPS equations (\ref{BPS1}) and (\ref{BPS2}) with initial values  $\phi _{1}^{k+}(0)=\phi _{2}^{k+}(0)=\phi _{2}^{k-}(0)=\phi
_{1}^{a-}(0)=\pi /2$ and $\phi _{1}^{a+}(0)=\phi _{2}^{a+}(0)=\phi
_{1}^{k-}(0)=\phi _{2}^{a-}(0)=-\pi /2$ for $\lambda=3 $, $\mu =0.5$. }
       \label{Fig1}}

We observe that the real parts are perturbed versions of the smooth kink and
antikink solution of the Hermitian case, which exhibit more and more
oscillations near the origin as $\lambda $ increases. Asymptotically the
solutions of the Hermitian and non-Hermitian cases tend to the same value.
Crucially, we read off the $\mathcal{CPT}$-symmetry (\ref{A}) for the
solutions 
\begin{equation}
\phi _{1}^{k\pm }(x)=\left[ \phi _{1}^{a\mp }(-x)\right] ^{\dagger },\quad
\phi _{2}^{k+}(x)=-\left[ \phi _{2}^{k-}(-x)\right] ^{\dagger },\quad \phi
_{2}^{a+}(x)=-\left[ \phi _{2}^{a-}(-x)\right] ^{\dagger },  \label{cpt}
\end{equation}%
from which we derive for the potential%
\begin{equation}
\mathcal{V}_{\lambda }\left[ \phi _{1}^{k+}(x),\phi _{2}^{k+}(x)\right] =%
\mathcal{V}_{\lambda }\left\{ \left[ \phi _{1}^{a-}(-x)\right] ^{\dagger },-%
\left[ \phi _{2}^{k-}(-x)\right] ^{\dagger }\right\} =\mathcal{V}_{\lambda
}^{\dagger }\left\{ \phi _{1}^{a-}(-x),\phi _{2}^{k-}(-x)\right\} .
\end{equation}%
This is once more the second option in (\ref{VV}). Thus assuming the
energies of kinks and antikinks in the $+$ system are the same as the
antikinks and kinks in the $-$ system, respectively, this energy is
guaranteed to be real.

Since the limits $x\rightarrow \pm \infty $ for these solutions are the same
as for $\lambda =0$, the expression for the energy $E(\mu )$ in (\ref{Emu})
holds for all values of $\lambda $. Considering the expression in (\ref{Emu}%
) it appears that $\mu =1$ is an exceptional point of the system and that
for $\left\vert \mu \right\vert >1$ one might obtain complex conjugate pairs
of eigenvalues. However, just as in the previous model, when the threshold
is passed into that region the asymptotic limits of the kink solutions are
no longer defined so that the expression for the energy becomes meaningless.
Moreover, when defining the theory on a finite interval in space the energy
is actually still real and does not occur in complex conjugate pairs. For an
exceptional point to emerge we would also expect that the antilinear $%
\mathcal{CPT}$-symmetry (\ref{cpt}) becomes broken when passing a genuine
exceptional point. However, this symmetry is still preserved in the regime $%
\left\vert \mu \right\vert >1$. Hence we conclude that $\mu =1$ is not an
exceptional point.

Next we identify the precise relation on which vacua are connected by which
of the various BPS solutions. The infinite amount of vacua of the potential (%
\ref{V}) are easily found to be%
\begin{equation}
v_{1}^{(n,m)}=(\arcsin \mu +2\pi n,m\pi ),\quad \text{and\quad }%
v_{2}^{(n,m)}=(\pi -\arcsin \mu +2n\pi ,m\pi ),
\end{equation}%
corresponding to the fixed points of the dynamical system (\ref{BPS1}) and (%
\ref{BPS2}), that are the solutions of $G_{1}^{\pm }(\phi _{1},\phi
_{2})=G_{2}^{\pm }(\phi _{1},\phi _{2})=0$. Computing once more the
eigenvalues of the Jacobian matrix at these fixed points%
\begin{equation}
J=\left. \left( 
\begin{array}{ll}
\partial _{\phi _{1}}G_{1}^{\pm } & \partial _{\phi _{2}}G_{1}^{\pm } \\ 
\partial _{\phi _{1}}G_{2}^{\pm } & \partial _{\phi _{2}}G_{2}^{\pm }%
\end{array}%
\right) \right\vert _{v_{j}^{(n,m)}},
\end{equation}%
with $j=1,2$, we find for the $+$ system that $J(v_{1}^{(n,2m)})$ has two
positive eigenvalues, $J(v_{2}^{(n,2m+1)})$ has two negative eigenvalues and 
$J(v_{1}^{(n,2m+1)})$, $J(v_{2}^{(n,2m)})$ have a positive and a negative
eigenvalue. For the $-$ system the signs are reversed. Thus the vacua $%
v_{1}^{(n,2m+1)}$, $v_{2}^{(n,2m)}$ are always saddle points, $%
v_{1}^{(n,2m)} $ are unstable/stable nodes ($G^{+}/G^{-}$) and $%
v_{2}^{(n,2m+1)}$ are stable/unstable nodes ($G^{-}/G^{+}$). Hence the kink
and antikink solutions only interpolate between the vacua $v_{1}^{(n,2m)}$
and $v_{2}^{(n,2m+1)}$ as indicated for an example in figure \ref{Fig2} with
the accompanying gradient flow.

\FIGURE{ \epsfig{file=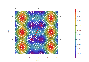, width=10cm} 
\caption{Real part of the potential $\mathcal{V}\left( \phi
_{1},\phi _{2}\right) $ as a function of $\func{Re}\phi _{1}$ and $\func{Re}\phi _{2}$ with the gradient flow of the real parts of $G^+$ superimposed in white.
The kink solutions $\phi _{1}^{k+}(x)$, $\phi _{2}^{k+}(x)$ interpolate
between the vacua $v_{1}^{(0,0)}$ and $v_{2}^{(0,1)}$ (red dots) as
indicated by the red solid trajectory.}
       \label{Fig2}}

The solutions depicted in figure \ref{Fig1} interpolate the vacua $%
v_{i}^{(n,m)}$ as%
\begin{equation}
v_{1}^{(0,0)}~\underrightarrow{\phi _{1}^{k+}\phi _{2}^{k+}}%
~v_{2}^{(0,1)},~~~v_{1}^{(0,0)}~\underrightarrow{\phi _{1}^{a+}\phi _{2}^{a+}%
}~v_{2}^{(-1,1)},~~~~~v_{1}^{(0,0)}\underrightarrow{\phi _{1}^{a-}\phi
_{2}^{k-}}~v_{2}^{(0,-1)},~~~~v_{1}^{(0,0)}~\underrightarrow{\phi
_{1}^{k-}\phi _{2}^{a-}}~v_{2}^{(-1,1)},
\end{equation}%
hence confirming the consistency of the above. The other vacua $v_{i}^{(n,m)}
$ for different choices of $n$ and $m$ are obtained by including the $n$%
-dependence into the solutions.

In both of our previous examples we have directly analyzed the complex
non-Hermitian systems. In analogy to the treatment of many quantum systems,
such an approach is especially meaningful under the assumption that there
exists an equivalent Hermitian system with the same energy. In the next
section we present such a system and thus further justify our approach.

\section{Complex extended sine-Gordon model and its Hermitian partner}

In this section we investigate a model with two complex fields consisting of
two copies of sine-Gordon models of which one is complex $\mathcal{PT}$%
-symmetrically extended 
\begin{equation}
\mathcal{V}\left( \phi _{1},\phi _{2}\right) =\frac{m^{2}}{2\mu ^{2}}\left[ 
\sqrt{1-\varepsilon ^{2}}-\cos (\mu \phi _{1})-i\varepsilon \sin (\mu \phi
_{1})\right] +\frac{m^{2}}{\mu ^{2}}\sin ^{2}\left( \frac{\mu }{2}\phi
_{2}\right) ,
\end{equation}%
with constants $m,\mu \in \mathbb{R}$ and $\left\vert \varepsilon
\right\vert \leq 1$. For simplicity we have not introduced an interaction
term between $\phi _{1}$ and $\phi _{2}$ as the feature we are trying to
illustrate can even be shown for a theory with one field only. We just keep
a second field to maintain a similarity with the previously discussed
systems and to allow for a direct comparison between the BPS solutions for
the two fields. The constant term proportional to $\sqrt{1-\varepsilon ^{2}}$
is introduced for convenience. In order to find a Hermitian partner
potential $\mathfrak{v}$ to the non-Hermitian potential $\mathcal{V}$ we
employ now a Dyson map originally found in \cite{Bender:2005hf}%
\begin{equation}
\tilde{\eta}=\exp \left[ \frac{\func{arctanh}\varepsilon }{\mu }\int dx\pi
_{1}(x,t)\right] .
\end{equation}%
Here the spacial momentum operator $\pi _{1}(x,t):=\partial _{t}\phi
_{1}(x,t)$ satisfies the canonical equal time commutation relation $\left[
\phi _{1}(x,t),\pi _{1}(y,t)\right] =i\delta (x-y)$. The inverse adjoint
action of $\tilde{\eta}$ on $\mathcal{V}$ then leads to 
\begin{equation}
\mathfrak{v}\left( \phi _{1},\phi _{2}\right) =\tilde{\eta}^{-1}\mathcal{V}%
\tilde{\eta}=\frac{m^{2}}{\mu ^{2}}\left[ \sqrt{1-\varepsilon ^{2}}\sin
^{2}\left( \frac{\mu }{2}\phi _{1}\right) +\sin ^{2}\left( \frac{\mu }{2}%
\phi _{2}\right) \right] ,
\end{equation}%
whereas the kinetic term remains unchanged as $\tilde{\eta}$ commutes with
it. Even though we are here mainly interested in the properties of classical
solutions, we briefly drew on the quantum field theory version of the model
in order to carry out the similarity transformation. The effect of the
adjoint action of $\tilde{\eta}$ on any smooth function of the fields $%
\left( \phi _{1},\phi _{2}\right) $ is $\left( \phi _{1},\phi _{2}\right)
\rightarrow \left( \phi _{1}+i/\mu \func{arctanh}\varepsilon ,\phi
_{2}\right) $.

We shall now demonstrate that the energies of the BPS solutions for the
system involving the non-Hermitian potential $\mathcal{V}$ and the Hermitian
potential $\mathfrak{v}$ are identical and real. Following the procedure of
the previous sections we first note that the potential $\mathcal{V}$ can be
derived from the pre-potential%
\begin{equation}
U\left( \phi _{1},\phi _{2}\right) =-\frac{2^{3/2}m}{\mu ^{2}}\left[ \left(
1-\varepsilon ^{2}\right) ^{1/4}\cos \left( \frac{\mu }{2}\phi _{1}-\frac{i}{%
2}\func{arctanh}\varepsilon \right) +\cos \left( \frac{\mu }{2}\phi
_{2}\right) \right] ,
\end{equation}%
when taking the metric of the target space simply to be diagonal $\limfunc{%
diag}\eta =(1,1)$. According to (\ref{BPS}) the two pairs of coupled BPS
equations are therefore%
\begin{eqnarray}
BPS_{1}^{\pm } &:&~~\partial _{x}\phi _{1}=\pm \!\frac{m\sqrt{2}\left(
1-\varepsilon ^{2}\right) ^{\frac{1}{4}}}{\mu }\!\sin \left( \frac{\mu \phi
_{1}}{2}-\frac{i}{2}\func{arctanh}\varepsilon \right) ,~~~~\   \label{BP1} \\
BPS_{2}^{\pm } &:&~~\partial _{x}\phi _{2}=\pm \frac{m\sqrt{2}}{\mu }\sin
\left( \frac{\mu }{2}\phi _{2}\right) .  \label{BP2}
\end{eqnarray}%
Once again we can identify a pair of modified $\mathcal{CPT}$%
-transformations under which these equations are compatible 
\begin{equation}
\mathcal{CPT}_{\pm }:~\phi _{1}(x)\rightarrow -\left[ \phi _{1}(-x)\right]
^{\dagger }\text{, ~~\ ~}\phi _{2}(x)\rightarrow \pm \left[ \phi _{2}(-x)%
\right] ^{\dagger },\text{\quad }\Leftrightarrow \quad BPS_{i}^{\pm
}\rightarrow \left( BPS_{i}^{\mp }\right) ^{\ast }.
\end{equation}%
We solve the equations (\ref{BP1}) and (\ref{BP2}) by%
\begin{eqnarray}
\phi _{1}^{k/a+}(x) &=&-\left[ \phi _{1}^{k/a-}(-x)\right] ^{\ast }=\pm 
\frac{4}{\mu }\arctan \left[ e^{mx\left( 1-\varepsilon ^{2}\right) ^{1/4}/%
\sqrt{2}+\mu \kappa _{1}/2}\right] +\frac{i}{\mu }\func{arctanh}\varepsilon ,
\label{phi1} \\
\phi _{2}^{k/a+}(x) &=&-\left[ \phi _{2}^{k/a-}(-x)\right] ^{\ast }=\pm 
\frac{4}{\mu }\arctan \left[ e^{mx/\sqrt{2}+\mu \kappa _{2}/2}\right] ,
\label{phi2}
\end{eqnarray}%
with integration constants $\kappa _{1},\kappa _{2}\in \mathbb{C}$. The
solution respect the $\mathcal{CPT}_{-}$-symmetry as indicated, which leads
to the relation%
\begin{equation}
\mathcal{V}\left[ \phi _{1}^{k/a+}(x),\phi _{2}^{k/a+}(x)\right] =\mathcal{V}%
^{\ast }\left[ \phi _{1}^{k/a-}(-x),\phi _{2}^{k/a-}(-x)\right] ,
\end{equation}%
for the potential that guarantees the reality of the energy when arguing
along the same lines as above.

We may of course also compute the energies directly from the asymptotic
limits of the solutions. For $\left\vert \varepsilon \right\vert \leq 1$ we
find 
\begin{eqnarray}
\lim_{x\rightarrow \pm \infty }\phi _{j}^{k+}(x) &=&\lim_{x\rightarrow \mp
\infty }\phi _{j}^{a-}(x)=\frac{\pi }{\mu }\pm \frac{\pi }{\mu }+\delta _{1j}%
\frac{i}{\mu }\func{arctanh}\varepsilon \text{, ~} \\
\lim_{x\rightarrow \pm \infty }\phi _{j}^{a+}(x) &=&\lim_{x\rightarrow \mp
\infty }\phi _{j}^{k-}(x)=-\frac{\pi }{\mu }\mp \frac{\pi }{\mu }+\delta
_{1j}\frac{i}{\mu }\func{arctanh}\varepsilon \text{,\ }
\end{eqnarray}%
which by (\ref{Q2}) gives the real energies%
\begin{equation}
E^{\phi _{1}^{pn},\phi _{2}^{qn}}(m,\mu ,\varepsilon )=\frac{4\sqrt{2}m}{\mu
^{2}}\left[ 1+\left( 1-\varepsilon ^{2}\right) ^{1/4}\right] ,~~\ \
p,q=k,a;~n=\pm ;~~~m,\mu \in \mathbb{R}  \label{E}
\end{equation}%
The special point $\varepsilon =1$ is not an exceptional point as the BPS
solutions for $\phi _{1}$ and $\phi _{2}$ have no definite asymptotic
values. For $\left\vert \varepsilon \right\vert >1$ the energies become
complex, albeit not complex conjugate. The reason for the latter is that the 
$\mathcal{CPT}$-symmetry is not just broken for the solutions, but also at
the level of the Hamiltonian. It is now easy to verify that the
pre-potential $\mathfrak{u}\left( \phi _{1},\phi _{2}\right) $ leading to
the real potential $\mathfrak{v}\left( \phi _{1},\phi _{2}\right) $ is
simply obtained as $\mathfrak{u}=\eta ^{-1}U\eta $. The solutions to the
real BPS equations are then given by (\ref{phi1}) and (\ref{phi2}) with $%
\left( \phi _{1},\phi _{2}\right) \rightarrow \left( \phi _{1}-i/\mu \func{%
arctanh}\varepsilon ,\phi _{2}\right) $. The expression for the energy $%
E=\lim_{x\rightarrow \infty }\mathfrak{u}\left( \phi _{1},\phi _{2}\right)
-\lim_{x\rightarrow -\infty }\mathfrak{u}\left( \phi _{1},\phi _{2}\right) $
is then the same as the one in (\ref{E}).

\section{Conclusions}

By assuming the non-Hermitian field theories to possess self-dual and
anti-self-dual fields we have derived their BPS soliton equations. We have
solved these equations for their complex kink and antikink solutions for
three different types of systems. We demonstrated that the solutions found
exhibit different types of modified antilinear $\mathcal{CPT}$-symmetries
relating the two versions of the BPS soliton equations. These symmetries
were shown to lead to real energies on general grounds in certain regimes of
the parameter space. For each of the systems we computed the topological
energy that saturates the Bogomolny bound confirming the generic result. We
observed that despite the fact that the BPS solutions are complex the
corresponding energies are real. Technically this is due to the fact that
contributions to the imaginary parts of the pre-potential are the same at
spacial plus and minus infinity. Crucially we found that the $\mathcal{CPT}$%
-symmetries can not be utilized directly on the self-dual part of the BPS
equation. However, taking both signs in (\ref{self}) into account the
symmetries can be identified.

For two of the systems we demonstrated explicitly how the kink/antikink
solutions interpolate between certain types of vacua corresponding always to
unstable fixed points at negative spacial infinity and stable fixed points
at positive spacial infinity. For the complex extended sine-Gordon model we
made the pseudo-Hermitian approach explicit and mapped the corresponding
non-Hermitian Hamiltonian to a Hermitian partner Hamiltonian by means of a
Dyson map. As the energies are preserved in this process and the Hermitian
theory always possesses real energies, this establishes the reality of the
energy computed from complex BPS solutions in the non-Hermitian theory.

There are evidently a number of interesting follow up problems and open
question. Since in none of the models we treated the transition point in
parameter space from real to complex or ill-defined energies led to a
genuine exceptional point, it remains an open question whether such type of
systems can be constructed. As our scheme is very general it should be
applicable to all non-Hermitian field theories that admit the described
self-dual and anti-self dual symmetries. Hence it would be interesting to
see the working of the above for more involved theories, possibly with a
larger field content. It would also be interesting to further compare with
an alternative approach to non-Hermitian field theories pursued in \cite%
{Kings1,Kings2,Kings3,Kings4,Kings5,Kings6} and investigate whether the BPS
soliton solutions derived in that framework also posses real energies.

%%\bibliographystyle{phreport}
%%\bibliography{ref.bib}

\end{document}